# *Perspective*

## ***Light-sheet microscopy in cancer pathology: current applications, clinical challenges and future perspectives***

*Uma Pisarović[1], Taichi Ochi[2,3], Ludovico Silvestri[4], Iryna Samarska[5,9], Thiemo J.A. van Nijnatten[6,9], Tom Marcelissen[7], Mari van Hout[5,9], Francesco Missale[8,9], Loes F.S. Kooreman[5,9], Axel zur Hausen[5,9], Anna Schueth[1,9,10].*

[1]Department of Genetics & Cell Biology, Onco-metabolism group, Maastricht University (UM), Maastricht, the Netherlands

[2]Department of Pharmacotherapy, Epidemiology & Economics, Groningen Research Institute of Pharmacy, University of Groningen, Groningen, The Netherlands, University of Groningen, Groningen, the Netherlands

[3]University Library, University of Groningen, Groningen, The Netherlands

[4]Department of Physics, LENS - European Laboratory for Non-Linear Spectroscopy, University of Florence

[5]Department of Pathology, Maastricht University Medical Centre (MUMC+), Maastricht, the Netherlands

[6]Department of Radiology and Nuclear Medicine, Maastricht University Medical Centre (Maastricht UMC+), Maastricht, the Netherlands

[7]Department of Urology, Maastricht University Medical Centre (MUMC+), Maastricht, the Netherlands

[8]Department of Otorhinolaryngology, Head & Neck Surgery, Maastricht University Medical Centre (MUMC+), Maastricht, the Netherlands

[9]GROW Research Institute for Oncology and Reproduction, UM, Maastricht, the Netherlands

[10]NUTRIM Research Institute for Nutrition and Translational Research, UM, Maastricht, the Netherlands

*To whom correspondence should be addressed: anna.schueth@maastrichtuniversity.nl*

## Abstract

Light-sheet fluorescence microscopy (LSFM) has emerged as a powerful research technology for three-dimensional (3D) visualization of biological tissues and is increasingly applied to human cancer specimens, including optically cleared biopsies and surgical resection samples. In this Perspective article, we review current applications of LSFM in oncology, and discuss how recent developments in microscopy, tissue preparation, labelling, and computational analysis are expanding the possibilities of 3D cancer pathology. Although LSFM has demonstrated the feasibility of large-volume imaging in selected human specimens, its application remains predominantly within research and translational settings and is not currently part of routine clinical pathology. We therefore distinguish demonstrated applications from potential future uses. We discuss how LSFM could eventually contribute to selected diagnostic and surgical workflows by providing 3D information on tumour architecture, spatial heterogeneity, metastatic dissemination, and surgical margins. These applications remain prospective and will require substantial technical, clinical, regulatory, computational, and workflow validation before clinical implementation can be considered. Ultimately, LSFM may complement rather than replace conventional histopathology by extending pathological assessment from two-dimensional tissue sections towards 3D spatial analysis.

### *Light-sheet fluorescence microscopy for large-scale three-dimensional cancer imaging*

Understanding cancerous tissue architecture and disease progression requires visualization of tumour organization across spatial scales, ranging from mesoscopic tissue overviews to cellular detailed regions of interest. Light-sheet fluorescence microscopy (LSFM) has emerged as a suitable imaging approach that enables three-dimensional (3D) visualization of intact cancerous specimens, such as human tissue samples at millimetre-scale in depths following optical tissue clearing protocol, as well as in vitro models and organoids [1-3]. By combining selective plane illumination with optical clearing strategies, LSFM enables rapid volumetric imaging of large biological specimens while minimizing photobleaching compared with point-scanning microscopy approaches [4, 5]. This unique combination of speed, depth, and volumetric coverage enables comprehensive analysis of the tumour architecture, including the spatial organization of malignant cells, stromal components, vasculature, and immune cell populations, while preserving morphological information across large tissue volumes.

Rather than representing a broadly established approach for cancer pathology, LSFM currently constitutes an emerging research technology that has demonstrated feasibility in selected human cancer specimens and experimental models. The principal opportunity lies in its ability to interrogate tissue volumes that are difficult to access with conventional section-based microscopy, while the extent to which this additional information can translate into clinically meaningful benefit remains an open question.

Confocal and two-photon microscopy (TPM) remain essential tools for high-resolution fluorescence imaging and provide excellent cellular and subcellular morphological detail [6-9]. However, their application to large intact specimen is limited by optical scattering and the reduced volumetric throughput of point-by-point scanning (<1 $mm^3$/day). Consequently, although these techniques are highly effective for resolving local cellular features, they are less suited for fast mapping of tissue-scale organization and the complex spatial interactions that define tumour heterogeneity [10]. Imaging millimetre-scale tissue volumes using confocal microscopy or TPM is therefore often impractical for routine analysis of large, optically heterogeneous, or refractive-index-mismatched specimens.

In contrast, LSFM enables rapid imaging of optically cleared tissues across substantially larger volumes, and intact clinical specimens, including FFPE core-needle biopsies and surgical resection samples [11-13]. Importantly, these applications represent research demonstrations in selected human specimens rather than established clinical pathology workflows. With LSFM, a single plane is imaged in a single shot, increasing imaging speed by several orders of magnitude [14]. Thanks to this paradigm change, LSFM allows imaging of large (cm-sized) 3D samples with high resolution, in reasonable time, and without the need for slicing the tissue [10, 15]. Importantly, when applied to histological specimen, LSFM needs to be coupled with clearing methods to make the tissue transparent and enable optical access to structures buried in the middle of the sample [16]. LSFM has been applied to analyse the 3D structure of multiple histological optically cleared samples derived from patients, ranging from 1 mm thick core-needle biopsies [17-20] to up to 7 $cm^3$ prostatectomy resection specimens [19]. By maintaining the native 3D context of tissues, LSFM provides access to architectural features that are difficult to capture using conventional section-based approaches, including tumour boundaries, invasive growth patterns, vascular networks, and heterogeneous cellular distributions. Beyond structural visualization, advances in multiplex fluorescence labelling and molecular imaging strategies are expanding LSFM from an anatomical imaging method toward a spatially resolved molecular profiling platform [21]. By combining volumetric imaging with immunofluorescence-based detection of specific cell populations, signalling pathways, and tumour-associated biomarkers, LSFM enables simultaneous characterization of tissue architecture and molecular heterogeneity throughout intact specimens. This integration of morphological, architectural, and molecular information provides new opportunities to investigate tumour evolution, cell–cell interactions, and disease-associated spatial patterns that remain difficult to resolve using conventional 2D approaches.

Histopathological diagnosis as the clinical gold standard relies primarily on hematoxylin and eosin (H&E)-stained sections approximately 3–5 µm thick examined by bright-field microscopy. This approach provides exceptional cellular resolution and remains highly effective for routine diagnosis. However, it represents only a limited sampling of the complete tissue volume. As a result, important 3D features, including spatial continuity of tumour structures, interactions between tumour and surrounding stroma, and the distribution of rare cellular populations, cannot be visualized.

LSFM-based volumetric imaging has the potential to complement conventional pathology by enabling visualization of tissue blocks of several millimetres in thickness, which are up to approximately 1000 times thicker than standard histological sections, while preserving native 3D organization. This capability is particularly relevant for clinically important processes in which spatial context is essential, including tumour invasion, vascular remodelling, lymphatic dissemination, and metastatic colonization of lymph nodes. Small or spatially discontinuous metastatic deposits that may be missed in limited 2D sections can be more comprehensively detected and interpreted within their surrounding tissue environment through volumetric imaging. However, the ability to visualize such features should not be equated with improved diagnostic sensitivity or clinical benefit until these endpoints have been systematically demonstrated in appropriately designed clinical studies. Although further optimization, standardization, and clinical validation are required, advances in high-speed and high-throughput LSFM have the potential to establish 3D imaging as a complementary approach to conventional histopathology. At present, LSFM is best regarded as an emerging research platform with potential relevance to future pathology rather than as an established diagnostic technology. By enabling simultaneous assessment of tumour morphology, tissue architecture, and molecular heterogeneity, LSFM may provide new insights into cancer biology and support future developments in spatial oncology and precision medicine.

### ***Light-sheet imaging application to human tissue samples***

To illustrate the potential of LSFM in cancer research, we present representative application cases in prostate and breast cancer. In addition, we address a few other cancer types where LSFM can be applied.

The described examples highlight how LSFM enables volumetric characterization of tumour morphology, tissue architecture, and spatial heterogeneity across clinically relevant specimens. **Table 1** summarizes current LSFM platforms applied to prostate and breast cancer models, including their technical features and biological applications. The choice of prostate and breast cancer reflects the availability of human-tissue studies rather than an indication that LSFM is more clinically mature in these cancers. The available evidence remains heterogeneous, encompassing different specimen types, fixation and clearing strategies, staining protocols, imaging modalities, and analytical approaches. These differences are important when considering the translation of experimental findings into clinical practice.

#### *Application case: Prostate cancer*

Recent studies have demonstrated the potential of LSFM as a transformative imaging approach for the analysis of prostate cancer specimens, particularly core-needle biopsies. By enabling rapid, high-resolution volumetric imaging of intact tissues, LSFM provides access to tumour features that are difficult to appreciate using conventional 2D histopathology. These developments are particularly relevant for prostate cancer, where diagnosis and risk stratification rely heavily on the assessment of glandular architecture and tumour morphology.

Glaser and colleagues developed an open-top light-sheet microscopy platform capable of high-throughput imaging of clinical core-needle prostate biopsies of approximately 1 mm thickness [5, 22]. Such biopsies are routinely obtained following elevated prostate-specific antigen (PSA) levels and remain central to clinical diagnosis. Current grading relies on the Gleason scoring system, which evaluates tumour differentiation based primarily on glandular architecture, ranging from well-formed glands (Gleason pattern 3) to highly disorganized and poorly differentiated structures (Gleason pattern 5). However, conventional histopathology is based on thin 3–5 μm H&E-stained sections, providing only a limited representation of the complex 3D organization of prostate tumours.

By contrast, LSFM enables volumetric visualization of entire biopsy specimens and larger prostatectomy samples, including tissue volumes ranging from several cubic millimetres to several cubic centimetres [19]. This provides a comprehensive assessment of tumour architecture, including glandular organization, tumour boundaries, and spatial heterogeneity throughout the specimen. Previous studies using confocal microscopy have demonstrated the feasibility of 3D visualization of architectural Gleason subgroups and gland formation in prostatectomy samples, achieving imaging depths of approximately 375 μm through analysis of 75 serial 5 μm sections [23]. This approach provided important insights into prostate tumour morphology and demonstrated that serial sectioning combined with confocal imaging can resolve 3D architectural patterns. However, the study was not primarily designed to achieve rapid, large-volume imaging of intact specimens. When the experimental or clinical question requires substantially larger tissue volumes to be surveyed at higher throughput, approaches such as LSFM offer important advantages by enabling direct volumetric imaging of optically cleared specimens without the need for extensive serial sectioning.

LSFM can reduce photobleaching during volumetric acquisition and enables direct imaging of intact, optically cleared specimens while preserving their 3D organization. For prostate cancer, this has been demonstrated in both core-needle biopsies and substantially larger prostatectomy specimens [19, 24]. The significance of these studies is not that they establish a new clinical diagnostic workflow, but that they demonstrate the technical feasibility of obtaining substantially more complete spatial information from human prostate specimens than is routinely available from a small number of histological sections.

One of the key advantages of LSFM is its ability to quantify tumour burden and spatial distribution throughout intact specimens. Small or discontinuous foci of adenocarcinoma that may be challenging to identify in individual 2D sections could potentially be detected more reliably through volumetric analysis. This represents a potential advantage of 3D imaging that warrants prospective clinical evaluation, particularly in heterogeneous tumours. However, whether 3D imaging consistently improves tumour detection, grading, staging, or patient management remains to be established. Furthermore, whole-organ imaging of optically cleared prostates may become feasible in the future, enabling investigation of tumour location and its spatial relationship with surrounding anatomical structures, including the urethra and prostatic capsule. Such information could be clinically valuable, as tumour proximity to the capsule is an important consideration in treatment planning, including decisions regarding nerve-sparing prostatectomy.

LSFM overcomes the limitation of 2D sections in pathology by enabling direct volumetric imaging of intact, optically cleared specimens with reduced photobleaching and preservation of tissue morphology. For example, complete visualization of cleared prostate cancer samples measuring approximately 40 × 35 × 5 mm has been achieved within several hours, corresponding to an imaging throughput of approximately 1.7 $cm^3$/h at 16.4 µm isotropic resolution, following a clearing procedure of approximately 10 days [19]. This example illustrates both the opportunity and the current translational challenge: although the imaging itself can be performed at substantial volumetric throughput, tissue preparation, clearing, staining, imaging, data processing, and interpretation constitute a substantially different workflow from routine histopathology.

At present, however, whole-organ LSFM of human prostate specimens should be regarded as a future research direction rather than a clinically available approach. Integration of LSFM-derived 3D histology with existing clinical imaging modalities may further enhance diagnostic and staging accuracy. For example, diffusion-weighted endorectal 3T magnetic resonance imaging (MRI) has been used to characterize prostate cancer prior to prostatectomy, with apparent diffusion coefficient (ADC) values correlating with tumour cell density [25]. Combining radiological imaging with volumetric histological information obtained through LSFM could provide complementary insights across spatial scales, linking macroscopic tumour localization with microscopic tissue organization. Such multimodal integration could eventually provide a framework for correlating radiological features with detailed tumour architecture, but prospective studies will be required to determine whether these correlations improve diagnosis, staging, treatment selection, or outcome prediction.

Beyond imaging alone, computational approaches are likely to play an important role in translating LSFM into clinically relevant applications. Recent work by Song and colleagues demonstrated that artificial intelligence (AI)-based analysis of LSFM datasets can achieve prostate cancer grading performance comparable to, or exceeding, conventional Gleason scoring based on H&E sections [26]. These findings are particularly important because they illustrate how computational analysis may help address the otherwise prohibitive volume of information generated by 3D pathology. Nevertheless, such studies remain research-stage demonstrations and require independent validation across institutions, scanners, tissue-processing protocols, and patient populations before clinical utility can be assessed. Together, these advances suggest that LSFM, particularly when combined with artificial intelligence and multimodal clinical imaging, has the potential to expand the capabilities of prostate cancer diagnosis by providing integrated information on tumour morphology, architecture, and spatial heterogeneity [26, 27].

It is important to consider that 3D prostate pathology does not necessarily need to replace the familiar 2D diagnostic paradigm. One possible future strategy could instead be to use the volumetric dataset to identify diagnostically informative regions or planes for subsequent pathologist review. Such an approach could preserve the interpretability and established expertise of conventional histopathology while exploiting the increased sampling provided by 3D imaging. Computational triage of large 3D datasets may therefore represent an important intermediate step between experimental volumetric imaging and eventual clinical application.

### *Application case: Breast cancer*

LSFM is gaining increasing attention in breast cancer research, with applications ranging from *in vitro* cell culture models and organoids to human surgical resection specimens. Breast cancer is the most diagnosed cancer among women, with a lifetime risk of diagnosis estimated at 1 in 20 women [28]. If current incidence trends continue, it is estimated that by 2050 there will be approximately 3.2 million new breast cancer cases and 1.1 million breast cancer-related deaths worldwide [29]. These trends highlight the importance of developing improved approaches to understand tumour heterogeneity, tumour–microenvironment (TME) interactions, and spatial features that influence diagnosis, prognosis, and treatment response. By enabling 3D visualization of intact tissues, LSFM provides a unique opportunity to investigate breast cancer architecture beyond the limitations of conventional 2D histopathology.

Recent studies have begun to address the limitations of conventional slide-based pathology for breast tissue analysis, highlighting LSFM as a powerful approach for investigating complex tissue features, including tumour architecture and tumour–microenvironment interactions (**Figure 1**). Considering the diverse molecular and histological subtypes of breast cancer, LSFM may provide additional spatial information to strengthen the understanding of tumour organization and disease progression [30]. By bridging experimental imaging approaches with clinically relevant specimens, LSFM may contribute to future strategies for improved characterization of breast cancer heterogeneity.

Importantly, LSFM enables rapid 3D imaging of fresh, non-cleared breast surgical specimens. Open-top light-sheet microscopy (OTLSM) approaches have achieved imaging speeds of approximately 1.5 $cm^2$/min [31] and 4.8 $cm^2$/min (12.5 sec/$cm^2$) [1], with minimal staining requirements (~2.5 minutes) [31]. These high-speed acquisitions allow efficient visualization of millimetre-scale tissue regions while preserving relevant morphological information. Such approaches facilitate assessment of tumour architecture, vascular structures, and surgical margins in freshly resected tissue samples [3]. These studies provide important proof-of-principle demonstrations that rapid optical imaging can be performed on selected fresh surgical specimens. However, they should not be interpreted as evidence that LSFM is currently used for intraoperative diagnosis or surgical guidance. To date, such applications remain experimental, and substantial work is required to determine whether imaging speed, tissue preparation, image interpretation, and diagnostic performance can meet the requirements of clinical practice.

The distinction between experimental feasibility and clinical implementation is particularly important for surgical applications. Intraoperative pathology operates under stringent constraints regarding turnaround time, specimen handling, diagnostic certainty, communication with the surgical team, and reproducibility. Demonstrating that a specimen can be imaged rapidly is therefore only one component of a potential future workflow. Prospective studies will be needed to determine whether LSFM-derived information can reliably identify clinically relevant tumour margins and, importantly, whether this information changes surgical decision-making or patient outcomes.

Beyond malignant tissue, LSFM has also been applied to investigate normal breast tissue organization and cellular phenotypes. 3D imaging of the terminal ductal lobular unit (TDLU) architecture has provided insights into the organization of healthy breast tissue and the structural changes that may accompany disease development [32]. These studies demonstrate the ability of LSFM to examine breast tissue across multiple spatial scales, from cellular organization to tissue-level architecture, providing a broader understanding of the structural context in which breast cancer develops.

Another promising application of LSFM in breast cancer is lymph node staging. Accurate identification and quantification of metastatic deposits within lymph nodes remain essential components of breast cancer management; however, conventional histopathology relies on limited 2D sampling that may incompletely represent heterogeneous metastatic involvement. Barner et al. demonstrated that multiresolution LSFM provides a non-destructive 3D approach for analysing cleared formalin-fixed paraffin-embedded (FFPE) lymph nodes from breast cancer patients [33]. Their study showed that conventional 2D histological assessment underestimated the size of lymph node metastases by an average of 19% compared with LSFM-derived measurements. This finding illustrates the type of spatial information that may be gained from 3D pathology, but it does not by itself establish improved staging accuracy or clinical benefit. Future studies should determine whether volumetric assessment changes the detection of metastases, pathological staging, treatment decisions, or patient outcomes.

Together, these studies demonstrate the potential of LSFM to extend breast cancer pathology from 2D morphological assessment toward 3D spatial characterization. By enabling visualization of tumour architecture, microenvironmental interactions, vascular organization, and metastatic dissemination, LSFM provides complementary information to conventional histopathology**.** Future integration of LSFM with molecular imaging approaches, artificial intelligence-based analysis, and established clinical workflows may further expand its role in breast cancer research and precision oncology.

*Building application cases in different tumour types*

While advances have been made in developing LSFM applications for prostate and breast cancer tissues, extending this approach to a broader range of tumour types remains an important next step. The limited application of LSFM across diverse cancers represents a current limitation of the field, as different tumour types exhibit distinct architectural features, microenvironmental interactions, and metastatic behaviours that require comprehensive spatial characterization. Expanding LSFM studies across additional cancer models will therefore be essential to determine its broader applicability in cancer pathology and to uncover tumour-specific biological processes.

Among the most common cancer types worldwide, lung and colorectal cancers represent important areas where improved understanding of tumour organization and progression could provide significant clinical and biological benefits [34]. Lung cancer accounts for nearly 20% of all European cancer deaths, with non-small cell lung carcinoma (NSCLC) comprising most lung cancer cases. Colorectal cancer represents the second highest incidence cancer type globally, further emphasizing the need for improved approaches to understand tumour development and progression [34]. Head and neck cancer, the sixth most common cancer worldwide, is projected to cost the global economy $535 billion in lost output between 2018 and 2030, with per-patient treatment costs reaching tens of thousands of euros in Europe (e.g., about €18,500 for oral cavity cancer in Italy) and over $130,000 for oropharyngeal cancer in the US.

Initial investigations using LSFM in these cancer types, including lung and colorectal cancer spheroid models, have demonstrated its potential for studying metastatic processes and tumour organization [35]. The improved optical capabilities of LSFM enabled detailed visualization and quantification of F-actin organization, allowing improved assessment of cellular architecture and necrotic regions within cancer tissues. These analyses revealed that certain cancer spheroids exhibited increased density compared with other cancer cell models, and LSFM provided the ability to visualize and quantify structural changes associated with tumour growth and progression. In head and neck cancer recent advances were applied to small samples of cleared tongue tissue and assessed with LSFM.

These examples also illustrate an important principle for interpreting the current literature: LSFM applications span substantially different biological systems, from engineered spheroids and organoids to human biopsies and resection specimens. Evidence obtained in one category should not automatically be extrapolated to another. Promising results obtained in experimental tumour models do not establish feasibility, diagnostic performance, or clinical utility in human tissue.

Although LSFM applications in these tumour types are currently less developed compared with prostate and breast cancer, these early findings highlight the potential of volumetric imaging to investigate cancer-specific biological processes that are difficult to resolve using conventional 2D approaches. Continued expansion of LSFM studies across diverse tumour lines will be important to establish the broader utility of this technology, particularly for understanding the spatial determinants of tumour invasion, metastasis, and interactions within the tumour microenvironment.

Future studies should therefore prioritize human specimens and clinically relevant questions rather than simply increasing the number of tumour models investigated. Comparative studies across tumour types could help determine where the additional spatial information provided by LSFM is most valuable, for example, in cancers characterized by infiltrative growth, complex glandular architecture, heterogeneous stromal organization, or spatially distributed metastatic disease.

### *Technological advances and future trends in light-sheet microscopy for cancer pathology*

LSFM offers substantial potential for advancing cancer research, diagnostics, and prognostic evaluation by enabling large-scale 3D visualization of intact tissues. However, the distinction between technological capability and clinical utility is central to the future development of the field. LSFM currently represents an emerging and increasingly powerful research technology in oncology, with demonstrated applications in selected human cancer specimens and experimental models, rather than an established clinical pathology platform. As the field continues to mature, expanding LSFM applications across diverse cancer types and clinically relevant specimens represents an important direction for future development. The major question for the coming years is therefore not simply whether LSFM can image larger volumes faster, but whether the additional information generated by 3D imaging can be converted into reproducible, clinically actionable information.

A critical step toward clinical adoption will be strengthening collaboration between LSFM researchers, pathologists, radiologists, and oncology clinicians. Close interaction between technology developers and clinical experts, particularly within hospital-adjacent research environments, will be essential for evaluating LSFM-based approaches using real patient samples and identifying clinical scenarios where 3D imaging provides meaningful additional information. As LSFM platforms continue to improve in speed, automation, and accessibility, adoption in preclinical and translational research settings is expected to increase.

Nevertheless, several challenges remain before LSFM can be incorporated into routine clinical practice. Compared with digital pathology technologies, LSFM currently remains at an earlier stage of clinical implementation. LSFM is currently not part of routine clinical pathology, and published applications involving human specimens have predominantly been performed in research settings. This distinction should remain explicit when considering the future clinical potential of the technology.

Digital pathology, also referred to as whole-slide imaging or digital slide technology, enables digitization of glass slides and supports improved diagnostic workflows through digital visualization, annotation, and data sharing [36]. Significant efforts have been undertaken to integrate digital pathology into clinical workflows, including practical implementation strategies and workflow optimization [37]. Following approval by the US Food and Drug Administration (FDA) for the use of digital pathology in primary diagnosis, adoption of this technology has continued to increase, although challenges remain regarding broader implementation and workflow integration [38].

Although digital pathology provides substantial benefits through improved slide interpretation, remote consultation, data sharing, and computational analysis, it remains fundamentally limited by the two-dimensional nature of conventional tissue sections. Consequently, important aspects of tumour architecture and microenvironmental organization may remain inaccessible. A recent comparison of missing tissue within breast tissue slides identified missing tissue frequencies ranging from 2% to 19% [39]. In many slides, missing tissue areas were less than 10% and were located peripherally, meaning that they did not alter the final diagnosis. However, even small loss of spatial information may become relevant when assessing heterogeneous tumours where critical biological features are unevenly distributed.

The comparison with digital pathology also highlights the magnitude of the translational challenge. Digital pathology was built around a workflow that retained familiar tissue processing and diagnostic interpretation while replacing the physical microscope with a digital representation of the histological slide. LSFM introduces a more fundamental change: tissue preparation, optical properties, staining, imaging geometry, data volume, visualization, and interpretation all differ from conventional pathology. Consequently, clinical translation will require more than demonstrating superior imaging performance as it will require development of an end-to-end workflow that can be reliably implemented within pathology practice.

To support implementation of LSFM, rigorous analytical and clinical validation against current diagnostic standards, particularly 2D H&E-based histopathology, remains essential. Although LSFM has demonstrated clear advantages for high-speed imaging of millimetre to centimetre-scale tissue specimens, validation of the complete workflow, including tissue clearing, staining protocols, image acquisition, and interpretation is required. Importantly, validation cannot be limited to technical agreement between LSFM images and conventional histology. Studies will need to determine diagnostic accuracy, interobserver reproducibility, failure modes, clinically relevant sensitivity and specificity, turnaround time, and the effect of LSFM-derived information on clinical decision-making. It must be established that individual optical sections generated by LSFM reliably correspond with conventional H&E sections in terms of diagnostically relevant information.

Beyond analytical validation, health technology assessments (HTAs) will be required to determine where LSFM can provide clinical value, including its position within diagnostic pathways, comparative cost-effectiveness, resource requirements, and potential impact on patient outcomes. Currently, the initial cost of LSFM systems remains a barrier to widespread adoption. However, continued technological development, increased availability of commercial "push-button" LSFM platforms, and the expansion of do-it-yourself (DIY) approaches supported by open-source hardware and software resources may reduce barriers to implementation. The emergence of a globally active LSFM community focused on open science and resource sharing further supports accessibility and continued innovation.

Clinical translation will also require consideration of tissue logistics. Conventional pathology has highly standardized pathways for specimen reception, fixation, grossing, sampling, processing, embedding, sectioning, staining, archiving, and reporting. An LSFM-based workflow would need to determine where tissue clearing and staining occur, how specimens are tracked, how long preparation can take, how tissue morphology is affected by processing, and how LSFM imaging relates to tissue that subsequently needs to be archived or subjected to molecular testing. These practical considerations may be as important for clinical adoption as the microscope itself.

Successful clinical implementation will also depend on addressing challenges associated with the management and interpretation of large-scale 3D datasets. Efficient visualization, processing, and analysis pipelines will require continued collaboration between computer scientists, imaging specialists, and pathologists to develop user-friendly platforms suitable for clinical environments. Automated analysis approaches will be particularly important for translating volumetric imaging data into clinically meaningful information. AI is expected to play an increasingly important role in this transition, as rapid advances in AI-based image analysis, including deep learning approaches and large language models, are creating new opportunities across oncology imaging applications. Recent studies have demonstrated the potential of AI-based analysis in prostate and breast cancer imaging, including applications for LSFM-derived datasets [40, 41]. For LSFM specifically, AI may become essential because the volume of information generated by intact-tissue imaging can exceed what a pathologist can reasonably inspect manually. Rather than replacing pathologist interpretation, computational methods could initially serve to identify regions of interest, quantify spatial features, or generate conventional-looking virtual sections from volumetric datasets. AI-powered interpretation of LSFM data may enhance diagnostic consistency and enable automated 3D histopathology, particularly for complex tumours where spatial heterogeneity limits conventional assessment.

***From research platform to potential clinical application***

The potential clinical applications of LSFM merit specific consideration, while recognizing that these applications remain prospective. Experimental studies have already demonstrated that LSFM can interrogate selected human specimens at spatial scales that are difficult to achieve with conventional histopathology. These demonstrations raise the possibility that the technology could eventually contribute to diagnostic workflows, margin assessment, tumour mapping, or other pathology applications. However, the existence of a technical capability does not establish clinical feasibility or utility.

One possible future application is the assessment of surgical specimens. Rapid imaging of freshly resected tissue has been explored experimentally, particularly using open-top configurations designed to image tissue surfaces or relatively shallow volumes. Such approaches are conceptually attractive for surgical margin assessment because they could provide spatial information without requiring immediate conventional sectioning. Nevertheless, an experimental demonstration of rapid imaging should not be equated with intraoperative pathology. A clinically useful system would need to meet stringent requirements for specimen handling, preparation time, image quality, diagnostic interpretation, reproducibility, and communication with the surgical team. Prospective clinical studies would ultimately be required to determine whether LSFM-derived information can reliably identify clinically relevant margins and whether its use improves surgical decision-making.

A second potential application is three-dimensional diagnostic pathology. Here, the major advantage of LSFM is not necessarily higher cellular resolution than conventional microscopy, but the ability to sample a much larger tissue volume while retaining spatial relationships. This could be particularly valuable in tumours with heterogeneous architecture or discontinuous disease. However, the clinical relevance of increased sampling remains to be demonstrated prospectively. A larger volume does not automatically produce a better diagnosis, and the interpretation of 3D datasets introduces its own challenges. A third possibility is the development of LSFM as a complementary research-to-clinic platform. Rather than replacing routine histopathology, LSFM could initially be used in selected cases where conventional sections leave clinically relevant spatial questions unresolved. Such a staged approach could provide an opportunity to evaluate the added value of 3D information while maintaining established diagnostic standards.

Finally, LSFM could provide an important bridge between pathology and other spatial imaging modalities. Integration with radiological imaging, molecular profiling, and computational pathology could enable correlation of tumour features across scales, from whole-organ or whole-specimen localization to cellular and molecular architecture. Such multimodal approaches may ultimately be more clinically informative than LSFM imaging alone. These applications should therefore be viewed as hypotheses for future translational research rather than established uses of LSFM. The central challenge for the field is to identify clinical questions for which the additional information provided by volumetric imaging is sufficiently valuable to justify the additional complexity of the workflow.

*Barriers to clinical translation*

Several barriers must be addressed before LSFM can move beyond research applications. The first is sample preparation. Optical clearing and deep-tissue labelling can require substantial processing time and may alter tissue dimensions, morphology, fluorescence, or compatibility with downstream assays. Different tissue types may respond differently to these procedures, making standardization across organs and tumour types challenging.

The second barrier is molecular labelling. Conventional H&E pathology provides a rapid and highly standardized representation of tissue morphology. In contrast, fluorescence-based LSFM generally requires defined labelling strategies, and penetration of antibodies or other probes through large tissue volumes can be heterogeneous. Multiplexing further increases experimental complexity. Development of robust, reproducible, clinically relevant labelling protocols will therefore be essential.

The third barrier is imaging and data management. Although LSFM can acquire large volumes rapidly, high-throughput imaging generates datasets that can be orders of magnitude larger than conventional whole-slide images. Storage, transfer, visualization, segmentation, registration, quality control, and long-term archiving therefore require dedicated computational infrastructure.

The fourth barrier is interpretation. Pathologists are trained primarily to interpret histological sections, and routine diagnostic criteria have evolved around 2D morphology. Translating established diagnostic concepts such as tumour grade, invasion, margin status, and metastatic involvement into 3D datasets will require both methodological development and potentially new diagnostic frameworks.

The fifth barrier is validation. Any future clinical application will require evidence that LSFM provides reproducible information across laboratories, instruments, tissue-processing protocols, and patient populations. Multicentre studies will be particularly important because many current demonstrations have been generated using specialized instruments and workflows developed within individual research laboratories.

Finally, clinical implementation will require evidence of clinical utility. A technically superior image is not necessarily a clinically useful test. Future studies should therefore move beyond demonstrations of image quality and ask whether LSFM changes diagnosis, grading, staging, treatment selection, surgical management, or patient outcomes. Cost, turnaround time, infrastructure requirements, training, and integration with existing pathology workflows will also need to be considered.

*Opportunities beyond diagnosis*

The potential value of LSFM extends beyond conventional diagnostic pathology. The ability to image intact tumour specimens provides an opportunity to study spatial relationships that may contribute to tumour progression, treatment resistance, and metastasis. For example, 3D characterization of tumour–stroma interactions, vascular organization, immune-cell distribution, and invasive fronts could generate new hypotheses regarding the mechanisms underlying tumour behaviour.

LSFM may also be particularly valuable for translational research because it can provide a bridge between experimental models and human tissue. Organoids, spheroids, xenografts, and patient-derived specimens can be investigated using related volumetric imaging principles, enabling comparison of tissue architecture across experimental systems. Such approaches could support drug development and biomarker discovery by revealing spatial phenotypes that are poorly captured by conventional 2D assays.

Another important opportunity is retrospective analysis of archived tissue. If robust clearing, labelling, and imaging methods can be standardized for FFPE material, existing pathology archives could become a resource for retrospective 3D studies. Large cohorts with long-term clinical follow-up could potentially be investigated for spatial biomarkers associated with recurrence, treatment response, or survival. This possibility is particularly attractive because it could allow the clinical relevance of 3D morphological features to be tested without requiring prospective collection of entirely new cohorts.

## ***Conclusion***

LSFM has emerged as a powerful imaging technology with diverse applications in cancer research and pathology. Its ability to visualize large-scale human tissue samples provides an opportunity to extend conventional histopathological analysis beyond 2D section-based assessment. Studies in prostate and breast cancer demonstrate the potential of LSFM to characterize tumour morphology, tissue architecture, metastatic dissemination, and TME interactions, providing complementary information that cannot be readily captured using conventional approaches.

Importantly, these applications remain heterogeneous and predominantly research based. LSFM is not currently part of routine clinical pathology, and no broad clinical workflow for either fixed or fresh human cancer specimens has been established. The existing literature instead demonstrates the feasibility of applying different LSFM approaches to selected specimen types and provides a foundation from which future clinical applications may be explored.

Beyond structural imaging, the integration of LSFM with molecular imaging approaches, including the visualization of RNA and proteins within intact tumour volumes, represents an important future direction for spatially resolved cancer pathology. Combining 3D anatomical information with molecular characterization may enable deeper insights into tumour biology and facilitate improved histopathological interpretation [42].

An additional opportunity lies in the application of LSFM to archived clinical specimens. The combination of deparaffinization, tissue clearing, and LSFM imaging of FFPE samples could enable retrospective 3D analysis of large patient cohorts, providing a valuable resource for translational cancer research. Such approaches may allow existing pathology archives to be revisited with spatial information that was previously inaccessible using conventional section-based methods. The development of these pipelines, including work from groups such as the Uhlén laboratory, highlights the potential of LSFM to contribute not only to future diagnostics but also to retrospective clinical studies and biomarker discovery.

Despite its promise, broad clinical adoption of LSFM will require continued technical validation, workflow standardization, and integration with established diagnostic pathways. Collaboration between LSFM researchers and clinical experts in pathology, radiology, and oncology will be essential to identify specific clinical questions for which 3D imaging provides meaningful additional value. Advances in automated image analysis, artificial intelligence, and data management will further be required to transform large-scale LSFM datasets into reproducible and clinically interpretable information.

The most important next step is therefore not simply to demonstrate increasingly large or faster LSFM datasets, but to determine where 3D information provides clinically meaningful advantages over existing approaches. This will require prospective and comparative studies, ultimately across multiple centres, evaluating diagnostic performance, reproducibility, workflow feasibility, cost, and clinical utility.

Ultimately, LSFM is unlikely to replace conventional histopathology but rather represents a complementary approach that expands pathology from two-dimensional morphology toward 3D spatial analysis. At its current stage, LSFM should be viewed as an emerging research technology with demonstrated applications in selected human cancer specimens and considerable potential for future translational development. Whether this potential can ultimately be realized in routine diagnostic or surgical workflows remains uncertain and will depend on overcoming substantial technical, biological, computational, clinical, regulatory, and organizational barriers.

## Figures

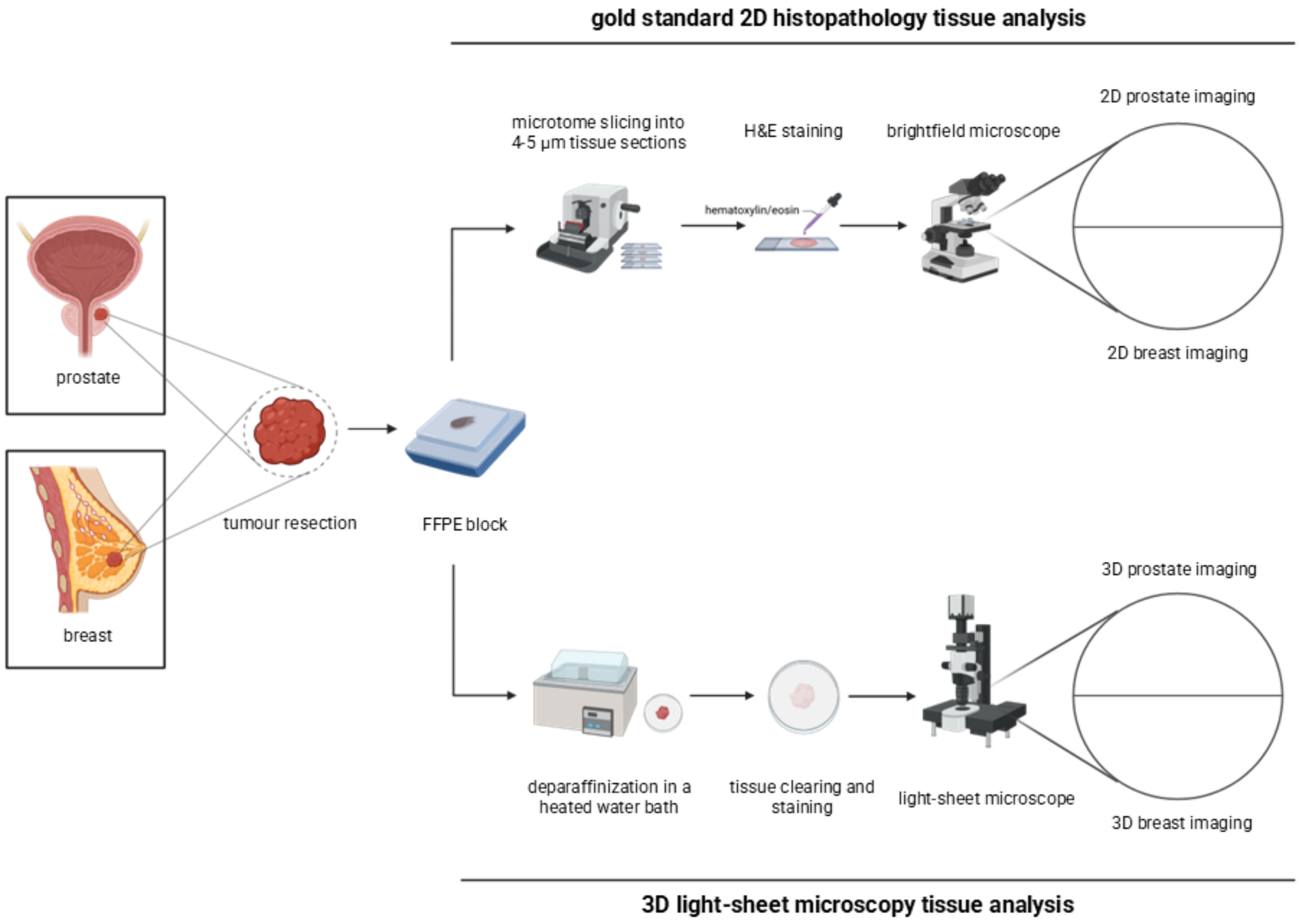

**Figure 1:** Light-sheet 3D assessment of large breast cancer samples vs. the 2D gold standard histopathology with thin tissue sections and bright-filed microscopy. (Figure created with Biorender)

## Tables

**Table 1:** Overview of light-sheet microscopy systems for different types of breast and prostate specimen.

| Organ | Sample description | Imaging modality |
|---|---|---|
| Breast | Fresh, not cleared cancer tissue mastectomy samples [1, 31] | Open-top light sheet microscope (OTLSM) |
| | Fixed, cleared cancer tissue mastectomy samples [3] | Ultramicroscope |
| | Fixed, cleared healthy tissue mastectomy samples [32] | Airy beam light-sheet microscopy |
| Prostate | Fresh, not cleared prostatectomy samples [1] | Open-top light sheet microscope (OTLSM) |
| | Fixed, cleared prostatectomy [19] | Open-top light sheet microscope (OTLSM)/ samples [33, 43], Cleared-tissue Dual view Illumination Microscope (ct-dSPIM) [19] |
| | Fresh, cleared core needle biopsy samples [1, 22] | Open-top light sheet microscope (OTLSM) |
| | Fixed, cleared core needle biopsy samples [24, 27, 44] | Open-top light sheet microscope (OTLSM) |

## Funding

This work has been supported by a Dutch Governmental Research grant and the University Fund Limburg (CoBes24.074) awarded to A.Schueth.

## Author contribution

All authors have written and contributed to the manuscript.

## Declaration of competing interest

T. van Nijnatten reports speaker honoraria, institutional grant support and medical advisory board meetings for Bayer and GE Healthcare, not related to the current manuscript. T. van Nijnatten reports medical advisory board for Screenpoint Medical, not related to the current manuscript.